\PassOptionsToPackage{hyphens}{url}
\PassOptionsToPackage{prologue,usenames,table,dvipsnames,x11names}{xcolor}
\documentclass[sigconf,screen,balance,natbib=false,authorversion]{acmart}
\usepackage{xspace}
\usepackage[abbreviations]{foreign}
\usepackage{siunitx}
\usepackage{graphicx}
\usepackage[
  backend=biber,
  style=ACM-Reference-Format,
  doi=false,
  isbn=false,
  minnames=3,
  maxnames=3,
]{biblatex}

\AtEveryBibitem{\clearfield{month}\clearlist{location}\clearfield{pagetotal}\clearfield{numpages}\ifentrytype{online}{}{\clearfield{url}}\ifentrytype{inproceedings}{\clearfield{pages}\clearfield{volume}\clearlist{publisher}\clearfield{articleno}}{}}

\renewbibmacro*{year}{\iffieldundef{year}{}{\printfield{year}}}

\usepackage{xpatch}

\makeatletter
\patchcmd\blx@bblinput{\blx@blxinit}
                      {\blx@blxinit
                       \def\jobname{paper}
                      }{}{\fail}
\makeatother

\def\bmPolybenchTwineWasmGlobslSlowdown{1.3}
 
 \def\bmSpeedtestOneTwineWasmGlobalSlowdown{2.7}
 \def\bmSpeedtestOneWatzWasmGlobalSlowdown{2.1}

\newcommand{\polybench}{PolyBench/C\xspace}

\newcommand{\wasm}{Wasm\xspace}

\makeatletter
\renewcommand{\@seccntformat}[1]{\csname the#1\endcsname\hspace{0.95em}} \makeatother

\copyrightyear{2022}
\acmYear{2022}
\setcopyright{acmcopyright}
\acmConference[FRAME '22]{Proceedings of the 2nd Workshop on Flexible Resource and Application Management on the Edge}{July 1, 2022}{Minneapolis, MN, USA}
\acmBooktitle{Proceedings of the 2nd Workshop on Flexible Resource and Application Management on the Edge (FRAME '22), July 1, 2022, Minneapolis, MN, USA}
\acmPrice{15.00}
\acmDOI{10.1145/3526059.3533618}
\acmISBN{978-1-4503-9310-2/22/07}



\AtBeginDocument{\providecommand\BibTeX{{\normalfont B\kern-0.5em{\scshape i\kern-0.25em b}\kern-0.8em\TeX}}}

\makeatletter
\begin{document}
\fancyhead{}

\title{\resizebox{0.98\linewidth}{!}{WebAssembly as a Common Layer for the Cloud-edge Continuum}}

\author{Jämes Ménétrey}
\email{james.menetrey@unine.ch}
\orcid{0000-0003-2470-2827}
\affiliation{\institution{University of Neuchâtel}
  \country{Switzerland}
}

\author{Marcelo Pasin}
\email{marcelo.pasin@he-arc.ch}
\orcid{0000-0002-3064-5315}
\affiliation{\institution{HES-SO University of Applied Sciences}
  \country{Switzerland}
}

\author{Pascal Felber}
\email{pascal.felber@unine.ch}
\orcid{0000-0003-1574-6721}
\affiliation{\institution{University of Neuchâtel}
  \country{Switzerland}
}

\author{Valerio Schiavoni}
\email{valerio.schiavoni@unine.ch}
\orcid{0000-0003-1493-6603}
\affiliation{\institution{University of Neuchâtel}
  \country{Switzerland}
}

\renewcommand{\shortauthors}{Ménétrey, et al.}

\begin{abstract}
    Over the last decade, the cloud computing landscape has transformed from a centralised architecture made of large data centres to a distributed and heterogeneous architecture embracing edge and IoT units.
    This shift has created the so-called \emph{cloud-edge continuum}, which closes the gap between large data centres and end-user devices.
    Existing solutions for programming the continuum are, however, dominated by proprietary silos and incompatible technologies, built around dedicated devices and run-time stacks.
    In this position paper, we motivate the need for an interoperable environment that would run seamlessly across hardware devices and software stacks, while achieving good performance and a high level of security --- a critical requirement when processing data off-premises.
    We argue that the technology provided by WebAssembly running on modern virtual machines and shielded within trusted execution environments, combined with a core set of services and support libraries, allows us to meet both goals.
    We also present preliminary results from a prototype deployed on the cloud-edge continuum.
\end{abstract}
 \maketitle

\section{Introduction}

In the last decades, numerous Web applications have been developed to be accessed from anywhere, including personal computers and smartphones.
Many of these programs were later moved to the cloud to be more practical or cheaper to maintain.
Other applications were initially designed for the cloud for scalability and availability, while relying on the cloud's naturally distributed and replicated nature.
No matter the reason, \emph{cloud computing} has become one of the main infrastructures supporting applications today.

The cloud was not always big as it is today.
It started simpler, with a handful of providers and basic services such as virtual machines and virtual storage.
Numerous cloud providers have come to exist to supply today's enormous cloud market demand.
Nowadays, some applications are built to exploit the cloud as a heterogeneous environment.
They can exploit it to obtain, for example, lower latency, more resilience, or legal compliance.
With a growing number of multi-cloud applications, dealing with different cloud providers and different cloud technologies has become a frequent issue.

Telecommunication companies started to deploy a more distributed infrastructure, with smaller, cloud-like clusters closer to the consumers of network-based services, in order to improve latency for their services.
Local administrations and other infrastructure providers such as energy and transportation followed suit, deploying small groups of rather powerful computing devices close to the human activity they support.
The use of these highly distributed devices has been collectively named \emph{edge computing}~\cite{7488250}.

To complete today's scenario, billions of sensing and actuating devices have been deployed, as what is called the \emph{Internet-of-things}, or IoT.
These devices are often tiny, with limited processing capabilities.
They execute a few simple tasks, such as sensing a temperature or turning on and off a light bulb.
Yet, they are connected to the Internet, often coordinating their function using some edge device and connecting users through a cloud service.

\begin{figure}[t]
    \centering
    \includegraphics[width=8cm]{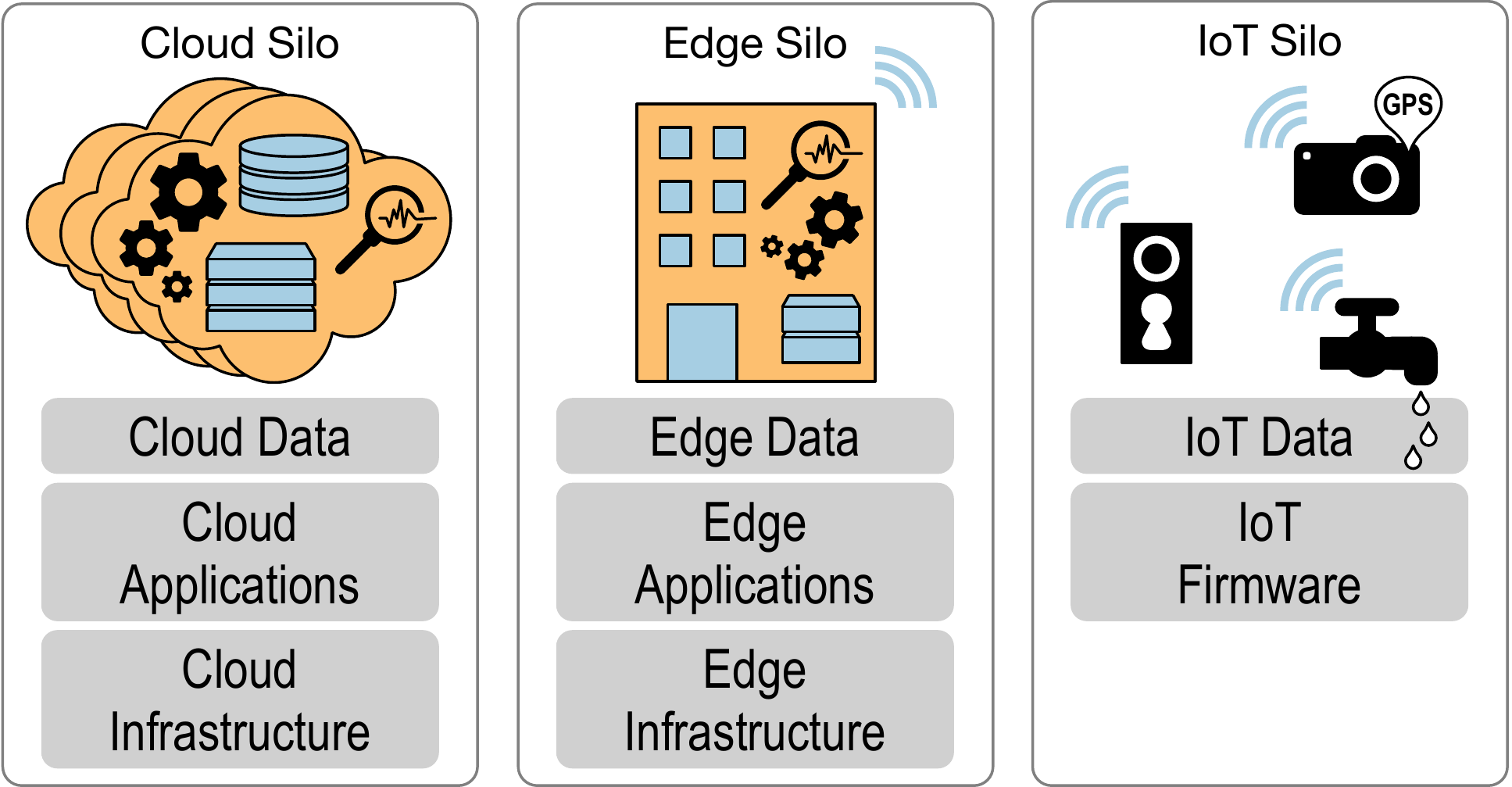}
    \caption{Independent cloud, edge, and IoT silos.}
    \vspace{-10pt}
    \label{fig:problemstatement}
\end{figure}

IoT, edge and cloud infrastructures form together what has been called the \emph{cloud-edge continuum}~\cite{bittencourt2018internet}.
This collective infrastructure is far from seamless today.
They actually exist in separate silos, dominated by proprietary solutions, as shown in Figure~\ref{fig:problemstatement}.
Developers of applications that span over the whole continuum must implement specific solutions for each silo, which are often built using incompatible software components.
The absence of a seamless environment makes it much harder to use the cloud-edge continuum.

Security has always been a component of applications shared among multiple users.
Traditional security used to deal with encryption, authentication and access control, and many established tools exist.
With the advent of the cloud, which is accessed through the Internet, security has become a fundamental component of all applications.
Providers, developers and users must be able to trust in the whole continuum --- cloud, edge and IoT --- in order to be sure that their data is safe and their computations are correct.

In this paper, we advocate that the technology provided by WebAssembly is adequate for implementing seamless applications across most hardware devices and software environments of the cloud-edge continuum, with the appropriate level of security.
We prove our claims by featuring a comparison of WebAssembly running benchmarks suites on two processor architectures.
To the best of our knowledge, this paper is the first to compare WebAssembly performance on different CPU architectures.
Modern hardware allows running WebAssembly while achieving good performance and a high level of security.
Furthermore, when paired with trusted computing, a technology that guarantees the confidentiality and integrity of secure applications, WebAssembly abstracts the complexity of software development while offering a trustworthy execution environment.
Nonetheless, many pieces are still missing from a full-fledged cloud-edge continuum.
Consequently, we also shed some light on the work yet to be covered by the research and industrial community.

In the following sections, we develop the current drawbacks of existing software architectures in more detail.
We then present WebAssembly and its advantages for executing applications in the cloud-edge continuum.
We complement our presentation with a preliminary performance comparison when executing selected applications using WebAssembly on two processor architectures to prove our claims of cloud to edge viability.
We conclude the paper with a few ideas for future work on the subject.
 \section{Building the cloud-edge continuum{\parfillskip=0pt}}

A typical cloud environment is a rather complex system, containing numerous (and different) hardware components.
Such components are exploited using extensive collections of software, managed by large engineering teams, and shared by many tenants.
Adding the edge (and IoT) to the picture pushes size and heterogeneity to another dimension.
An ideal seamless cloud-edge continuum should offer a lightweight execution environment with a similar (or even identical) software and hardware interface, allowing unmodified code to be executed in any machine in the system.

Some initiatives already exist for a common environment for cloud, edge and IoT silos.
The Java Virtual Machine (JVM) \cite{venners1998java} is one of the first practical common environment implementations that address the issue of having applications running on heterogeneous underlying systems.
To a considerable extent, the JVM is today one of the most comprehensive choices, with implementations for commodity servers to embedded devices.
Still, the JVM supports very few programming languages and adds substantial performance overheads compared to the native execution of C programs.
Java programs depend on a vast number of class libraries, imposing a large memory footprint for executing even the simplest of programs.
Containers appeared more recently \cite{271784}, as an alternative for running applications in heterogeneous environments.
Still, containers are defined for specific architectures and a particular operating system interface.
One needs to rely on recompilation to obtain containers that can run, for instance, on Intel and Arm devices (respectively popular as cloud and edge devices).
WebAssembly has the generality of JVM and the lightness of containers, allowing to build multi-platform software that can execute with negligible performance losses and small memory footprints.

To automatically deploy applications in a distributed system, one has to deal with aspects such as admission control and resource management, as well as monitoring and optimising the use of the devices to compute and communicate.
We are unaware of any practical, specific tool spanning over the whole cloud-edge continuum, but we assume it would be straightforward to adapt many of the existing tools developed for the cloud~\cite{costa2022orchestration}, provided that the underlying systems become more homogeneous.
Also, a few authors have already started working on models for integrating cloud and edge devices into a seamless system~\cite{bittencourt2018internet, balouek2019towards, 10.1145/3226644}.
We do not address the issue in this paper.
Instead, we propose to use a homogeneous runtime model to close the gap at low-level.

As said, security has become an essential issue in cloud systems.
Application users need guarantees that their data's confidentiality and integrity are respected.
These guarantees are hard to provide in a multi-tenant system, where co-tenants may abuse the system's vulnerabilities to uncover (or infer) someone else's application data.
It is even more complex when the infrastructure provider is curious, as it has all the administrative power needed to inspect all contents in all physical machines.
On the other hand, infrastructure providers wish to be protected from malicious tenants, who may want to exploit the infrastructure vulnerabilities for their own profit.

Edge computers are much more distributed when compared to the cloud.
They are installed in user buildings, shared infrastructures, or even next to roads, being impossible to maintain physical control over the resources.
Edge administrators have physical access to the edge devices they manage, with similar powers to cloud providers.
On the other hand, users are in the proximity of the edge devices and may even physically abuse them.
Edge-based infrastructure offers far fewer guarantees than the cloud.

Most recent versions of popular computer architectures include some form of a \emph{trusted execution environment} (TEE), a practical solution for establishing trust.
Such TEEs allow code execution in a segregated hardware section, where access is architecturally impossible from other software.
A TEE can execute a program and protect its data so that a machine administrator cannot access it.
Current hardware implementations may include an extra execution mode in the processor, or even memory encryption for TEE data.
The most popular implementation of TEE today is Intel's Secure Guard Extensions~(SGX)~\cite{mckeen2013innovative}, for which commercial cloud services such as Azure Confidential Computing~\cite{msazureconfidential} already exist.
A similar solution is necessary for edge deployments as well, where the most popular architecture is Arm, which in turn offers TrustZone~\cite{trustzone} as a TEE.
The existence of proprietary and incompatible solutions in the underlying hardware makes it harder to reuse trusted software from cloud to edge and vice versa.

\emph{Confidential containers} could be a practical alternative for deploying applications on the cloud-edge continuum, as proposed by Scontain~\cite{arnautov2016scone}.
They are similar to traditional containers, except they run entirely inside a trusted environment.
However, like other containers, they are platform dependent.
Also, they are costly in terms of the resources needed in many cases, as they may incorporate substantial amounts of operating system features.

Microsoft's Azure Sphere~\cite{azuresphere} follows the same idea but offers a unified programming model and support for certain trusted execution technologies.
As a proprietary solution, it is heavily dependent on other Microsoft services.
It offers a high-level interface but only supports a few programming languages.

By proposing WebAssembly as the execution model combined with trusted execution environments, we can offer a seamless portability base for running trusted applications.
The same base can be used to deploy applications on edge or cloud devices, with similar security guarantees.
Besides, it has been shown that a WebAssembly TEE enables a double-sided sandbox~\cite{goltzsche2019acctee} providing better security for the provider and the tenants.

Many different IoT infrastructures have been deployed and continuously generate data, feeding cloud applications worldwide.
Components in the application chains (IoT to edge to cloud) may be updated independently, adding new functionalities and removing vulnerabilities.
Particularly in this application area, we observe the growing use of federated machine learning, where edge devices collaborate to build a model without revealing all details of each user's data, helping to maintain data privacy.
Besides, attestation~\cite{menetrey2022} plays a fundamental role in such a dynamic, distributed scenario.
It allows for establishing trust in specific pieces of software, verifying their authenticity and integrity.
Through remote attestation, one can ensure to be communicating with a specific, trusted (attested) program remotely.
We believe that attestation plays an essential role in building a fully trusted environment for running cloud-edge continuum applications.
 \section{WebAssembly under the hood}

This section describes how the cloud-edge continuum can leverage WebAssembly as a unifying technology and the key benefits over the current state-of-the-art solutions.

\subsection{WebAssembly binary instruction format}

WebAssembly~(\wasm)~\cite{haas2017bringing, WebAssemblyCoreSpecification2} is a novel and general-purpose virtual instruction set architecture (ISA).
In contrast to previous efforts for platform-independent execution, such as Java from Oracle and Microsoft .NET, \wasm is developed by a consortium of technology companies from the beginning, such as Microsoft, Google and Mozilla, among others.
While it was initially designed to increase the performance of Web applications, \wasm does not depend on any Web-related features and is increasingly used for building standalone applications.
\wasm has many advantages to be used as a unified execution unit for the cloud-edge continuum.
First, \wasm is a compilation target for a wide variety of programming languages, enabling developers to write applications using their favourite programming languages and deploy them across the continuum without adaptation.
Second, contrary to Java and .NET, \wasm is compact, has minimal dependencies, and offers additional security benefits, such as sandboxing.

\wasm interacts with the underlying operating system thanks to the WebAssembly System Interface~(WASI)~\cite{Mozilla2019StandardizingWASI}, a specification standardising a POSIX-like interface.
WASI has been designed with conciseness and portability, enabling the platforms to implement the specifications with ease, ideal for constrained environments, such as IoT and edge devices, as well as TEEs.
It currently offers a set of 46 functions, allowing applications to interact with files, networking, and many other operating system functions.
Popular compilers for languages such as C and Rust seamlessly translate POSIX calls to WASI calls.
Additionally, WASI follows the concept of capability-based security, a security model where each resource (\eg socket, file) access must be granted by the \wasm runtime, enabling establishing a sandbox.
For example, WASI restricts the application to a subtree of the file system by introducing an abstraction layer between the \wasm program and the operating system interface.

Finally, \wasm runtimes can have different memory footprints depending on the execution model used, such as interpretation, just-in-time compilation (JIT), or ahead-of-time compilation (AOT).
As an example, WAMR~\cite{wamr} is a micro runtime aimed for edge devices that has a file size of \qty{209}{\kibi\byte} when running AOT code and \qty{230}{\kibi\byte} when interpreting, and \qty{41}{\mebi\byte} when executing JIT code.
A growing list of toolchains already support \wasm as a compilation target for different source languages, including C, C++ and Rust.
Examples are LLVM~\cite{lattner2004llvm}, an entire compilation infrastructure, and Emscripten~\cite{zakai2011emscripten}, a source-to-source compiler.
Support for other programming languages, including C\#, Go, Kotlin, Swift, and more are under active development.
For all these reasons, we believe \wasm to be an excellent choice for the binary architecture of the entire cloud-edge continuum.

\subsection{Trusted execution environments}

Trusted execution environments aim to provide safe and trustworthy code execution on (remote) untrusted hardware.
Hardware manufacturers have provided TEE implementations more than a decade ago, each one of them offering different features and guarantees.
The most influential TEEs that are currently marketed are Intel SGX~\cite{costan2016intel}, Arm TrustZone~\cite{trustzone}, and AMD Secure Encrypted Virtualization~(SEV)~\cite{sev}.
These technologies enable the processing of data, contained in isolated memory areas that cannot be accessed nor tampered with by more privileged software, such as the operating system or the hypervisor.
Hence, cloud providers and edge device owners with management rights or even physical control cannot access the data and computation of a tenant, protecting the confidentiality and integrity of their applications.

Cloud providers, such as Microsoft Azure and Google Cloud, already market confidential computing, and we expect widespread adoption of these services due to the demand driven by the cloud-edge continuum~\cite{msazureconfidential,googleconfidential}.
We observe that the rich ecosystem of trusted environments largely varies in terms of security, threat models, and implementation.
However, defining a common basis for trusted execution and making it widely available in both cloud and edge environments is an essential need for the continuum, and for the industry in general.
For that reason, Arm, Intel, Microsoft and others created the Confidential Computing Consortium~(CCC)~\cite{ccc}, supporting open-source projects for trusted execution technology under the umbrella of the Linux Foundation.
A unified abstraction for TEEs in the cloud-edge continuum must take support and shape from such ongoing efforts.
For that reason, the CCC is involved in many projects such as Enarx~\cite{enarx} and Veracruz~\cite{veracruz}, which aim to provide \wasm support in TEEs, independently from hardware.

In our previous work, we proposed a few solutions to execute general-purpose \wasm applications within TEEs.
We developed \textsc{Twine}~\cite{twine} to bring a \wasm runtime into Intel SGX enclaves, leveraging WASI to interact with the TEE facilities and the untrusted operating system.
More recently, we proposed \textsc{WaTZ}~\cite{watz}, a trusted runtime for Arm TrustZone with added remote attestation.
The latter, an essential feature for providing trust for remote applications, is surprisingly missing in Arm's architecture.
We believe that industrial versions of our prototypes will help paving the way to build distributed applications on the cloud-edge continuum that providers, developers and users can safely trust.

\subsection{Live migration of applications}

Migration is another need of the cloud-edge continuum that can be helped with the homogeneity offered by \wasm.
Migration may be needed for a variety of reasons.
Some applications have strict latency constraints, and may need to migrate to keep close to mobile users.
Some applications temporarily require high processing power and may need to migrate to powerful processors in the clouds.
Some applications may also move closer to where data is collected, because of legal regulations, because users want to process their private data locally, or simply because it is faster to process it closer to their sources.
Migration is much more of a challenge if the underlying environment is as heterogeneous as the continuum.

Previous work covered the needs and solutions to hand off virtual machines~\cite{10.1145/3132211.3134453}, requiring transferring large amounts of data representing memory or even disk images.
Transferring Intel SGX enclaves~\cite{8416483} not only is bound to a specific TEE technology, but also depends on the application help to provide its state, since the operating system cannot access enclave memory.
In contrast, \wasm offers a great environment to migrate running applications, thanks to its linear memory design and sandboxing mechanism.
Indeed, \wasm's memory is stored in a contiguous memory segment, where references in code are relative to its starting address.
Moving the linear memory to another address or machine does not involve any changes to the references in the application state.
Furthermore, the opened resources are tracked by the \wasm runtime with WASI, enabling the reproduction of external dependencies.
Jeong et al.~\cite{10.1145/3357223.3362735} studied the live migration of \wasm applications loaded in browsers by replicating the linear memory.
Nonetheless, migrations of \wasm applications with opened dependencies from cloud to edge machines are yet to be demonstrated.
The process of migrating executing software depends on the underlying system and its dependencies (\eg opened files, sockets).
We believe future work will enable applications to be seamlessly deployed and moved across the cloud-edge continuum, regardless of the devices' processor architecture, TEE technologies and operating systems.

\subsection{Technological limitations and pitfalls}

While we presented many advantages of \wasm, limitations also exist.
One first challenge is simply compiling applications in \wasm.
Even though compilers are mature enough to translate source code into \wasm bytecode (\eg LLVM), the support offered by WASI for surfacing system calls remains limited.
Lifting this limitation by extending WASI to match POSIX fully would probably restrict the ability to execute \wasm applications in various environments, such as Web browsers or TEEs.
Also, the behaviour of WASI diverges from POSIX by adding sandboxing.
One alternative is to avoid using WASI, as done by Emscripten, which compiles and defines the import section of \wasm applications with POSIX functions and system calls directly.
While this helps running legacy \wasm programs with only a few changes on POSIX systems, it reduces portability for platforms without POSIX, such as other OSes (\eg Windows) and restricted environments (\eg TEEs or IoT devices).

Executing \wasm code imposes a performance overhead, as it happens with all intermediary representations.
As we demonstrate in Section~\ref{sec:perf}, \wasm programs may be up to 3$\times$ slower when compared to their native version, depending on the type of workload.
This is explained by many factors, such as increased register pressure, additional branch instructions, increased code size, stack overflow checks and indirect call checks.
While some of these issues can be compensated by allowing compilers to spend more time generating better code, other factors are a consequence of the design constraints of \wasm, which would require changes in the \wasm specifications to be solved at the cost of complicating the implementation of the compilers and the way \wasm operates~\cite{234914}.

Another less perceptible limitation is the size of the allocatable volatile memory for \wasm applications.
\wasm uses a linear memory to store the heap of an executing program.
The linear memory is measured in pages, where each page has \qty{65536}{\byte} ($2^{16}$).
A linear memory instance can contain up to \num{65536} pages, for a total of \qty{4}{\gibi\byte} ($2^{32}$).
Besides, \wasm memory instructions' indices are 32-bit unsigned integers.
While most software does not require more than \qty{4}{\gibi\byte} of linear memory, this may restrict some applications of \wasm, such as training sizeable deep learning models or keeping large databases in memory.
Fortunately, recent proposals~\cite{memory64} aim to extend this limitation by increasing the number of allocatable pages to $2^{48}$, pushing the theoretical memory cap to \qty{16}{\exbi\byte} ($2^{64}$).

Finally, \wasm is still young and improving through community proposals.
Its second major version has been recently released~\cite{WebAssemblyCoreSpecification2}, introducing many features, such as reference types, bulk memory and SIMD instructions.
Future contributors may propose \wasm and WASI extensions to relax the limitations or extend the capabilities of the specifications.
For example, \emph{wasi-nn} is a proposal to add a WASI module for machine learning to facilitate model inferences~\cite{wasi-nn}.
In conclusion, we expect that the current limitations of \wasm will recede with respect to software compilation and deployment for the cloud-edge continuum, thanks to advances in compilation toolchains, extensions of the specifications, and better support of \wasm for all types of devices and environments, including TEEs.
 \begin{figure*}[t]
    \centering
    \includegraphics{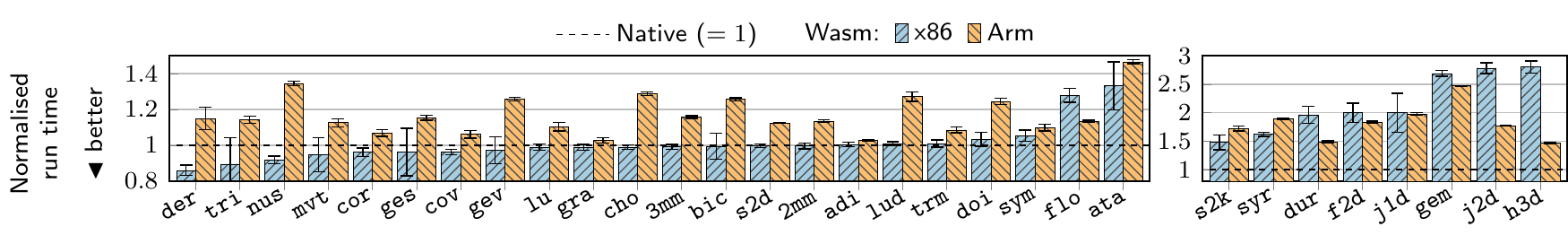}
    \vspace{-25pt}
    \caption{Relative performance of \texttt{Polybench/C} benchmarks.}
    \label{fig:polybench}
\end{figure*}

\begin{figure*}[t]
    \centering
    \includegraphics{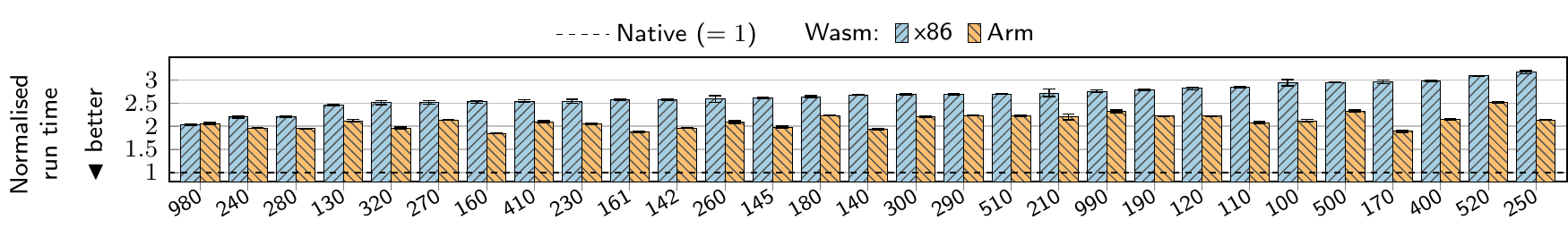}
    \vspace{-23pt}
    \caption{Relative performance of SQLite \texttt{Speedtest1} benchmarks.}
    \label{fig:speedtest}
\end{figure*}

\section{Performance}
\label{sec:perf}

This section shows how \wasm scales across the continuum, going from cloud to edge devices using different hardware configurations.
While recent research demonstrates the performance of \wasm in general, we are the first to compare two benchmark suites on different processor architectures.
Our goal here is to show that \wasm is viable for running code on a variety of devices for similar workloads, without suffering from significant performance overheads.

As an example of cloud server, we use a Supermicro 5019S-M2, equipped with an Intel Xeon E3-1275 v6 (\qty{3.8}{\giga\hertz}).
We settled for an off-the-shelf NXP MCIMX8M board as an edge device, equipped with an Arm Cortex-A53 (\qty{1.5}{\giga\hertz}).
While TEEs are not demonstrated in this work, these two platforms support trusted execution, namely Intel SGX for the former and Arm TrustZone for the latter.
We already illustrated how \wasm could be embedded within SGX and TrustZone in our previous work~\cite{twine,watz}.

We opted for WAMR as a \wasm runtime for its small size and portability across operating systems and constrained environments, such as TEEs.
The \wasm benchmarks are compiled into \wasm format using Clang, then compiled again ahead-of-time into a native format using the compiler provided by WAMR (\ie, \texttt{wamrc}).
Time is measured using the POSIX function \texttt{clock} in all the benchmarks and averaged using the median.

\subsection{\polybench micro-benchmarks}\label{sec:polybench}

\polybench\cite{polybench} is a CPU-bound benchmark suite comprised of various mathematical experiments and commonly used to evaluate the performance of \wasm applications and runtimes~\cite{234914,goltzsche2019acctee,twine}.
The name of the experiments have been abbreviated in this paper for conciseness.
We assessed 30 \polybench experiments and compared the performance overheads \wasm introduced on x86 and Arm architectures, relative to each of their native versions (plain x86-64 and Arm ELF binaries).
As such, these measurements compare how \wasm applications perform depending on the deployment target (\ie cloud or edge machines).

In Figure~\ref{fig:polybench}, we observe a similar slowdown between the \wasm experiments on both architectures.
Indeed, \wasm on x86 and Arm architectures both achieve a slowdown relative to native of \bmPolybenchTwineWasmGlobslSlowdown$\times$.
We identify and summarise the following groups based on the test performance results: 
\emph{(1)} the run time of \wasm and native are similar (\eg \texttt{lu}, \texttt{gra}, \texttt{adi}),
\emph{(2)} the run time of \wasm are similar, but slower than native (\eg \texttt{s2k}, \texttt{f2d}, \texttt{j1d})
\emph{(3)} the run time of \wasm is faster than native (\eg \texttt{der}, \texttt{tri}, \texttt{nus}), and
\emph{(4)} the run time of x86 \wasm are significantly slower than Arm \wasm and native (\eg \texttt{j2d}, \texttt{h3d}).

\wasm is naturally slower than native because of the increasing of register pressure and code size and the presence of extra branch statements, as discussed in previous work~\cite{234914}.
In some rare cases, \wasm may be faster than native thanks to a reduced number of cache misses, as we observed in our previous work~\cite{twine}.
Finally, some workloads are not well optimised when compiled in \wasm and then recompiled ahead-of-time into native code, as can be observed for the Arm versions on the left-hand side of Figure~\ref{fig:polybench}.

\subsection{SQLite macro-benchmarks}\label{sec:speedtest1}

SQLite~\cite{5231398} is a widely used full-fledged embeddable database.
Thanks to WASI, we showcase the versatility of \wasm by compiling and running SQLite outside of a browser.
As such, we diverted the operating system calls made by the database engine to be handled by WAMR.
For this purpose, we implemented a shim OS layer as the minimum set of POSIX functions in WASI necessary to support SQLite's in-memory databases.
We assessed SQLite performance using its official benchmark suite (\texttt{Speedtest1}~\cite{speedtest1}), running 29 out of the available 32 tests, covering a large spectrum of scenarios (we excluded 3 experiments because of issues with our shim layer).
Each \texttt{Speedtest1} experiment targets a single aspect of the database, \eg, selection using joins, or the update of indexed records.

Figure~\ref{fig:speedtest} presents our evaluation, where we compare the execution speed of \wasm, using both x86 and Arm architectures, and again normalise against the native run time.
Overall, the \wasm slowdown relative to native is \bmSpeedtestOneTwineWasmGlobalSlowdown$\times$ for x86 and \bmSpeedtestOneWatzWasmGlobalSlowdown$\times$ for Arm.
Most of the experiments located on the right-hand side of the figure, which are slower, are related to inserting, updating or deleting data (\ie 100-120, 180, 190, 230-250, 270-300, 400, 500).
The remaining experiments are related to data reading (\ie 130-145, 160, 161, 260, 320, 410, 510, 520) and housekeeping (\ie 980, 990).
Therefore, we correlate an increasing impact on the performance of write-intensive operations.
Finally, when preparing the experiments for this paper, we noticed a performance improvement in WAMR's ahead-of-time compiler compared to our results from 2020.
Indeed, in our previous work on \textsc{Twine}~\cite{twine}, we measured similar native performance on the same x86 hardware and a considerable worse performance when using WAMR (was 4.1$\times$).
This strengthens the perspective of using \wasm as a universal, lightweight, yet versatile bytecode to enable platform independence across the continuum.
 \section{Conclusion}

We envision the cloud-edge continuum as an interoperable, scalable and distributed system, where software may be located to any peer, regardless of the underlying platform.
This practice will transform the development lifecycle of future applications, enabling developers to focus on the business value instead of dealing with the complexity of each different piece of infrastructure.
\wasm is a perfect fit for that task, thanks to its abstraction from the operating system, device type, programming language, and the added security guarantees it can provide using TEEs.

We presented some performance measurements showing that \wasm is a viable alternative to native execution, with acceptable overheads.
We covered many aspects of how \wasm can be successfully adopted for the cloud-edge continuum, such as trusted computing, which enforces the applications' confidentiality and integrity, and live migrations, diminishing the latency or increasing the computation power by relocating running software seamlessly.

Remaining challenges concern enhancing the interoperability with the existing programming languages towards \wasm, while extending WASI to increase the capabilities of hosted applications.
For example, recent initiatives are bringing neural networks~\cite{wasi-nn} and parallelisation~\cite{wasi-parallel} into the WASI specifications.
Building middleware software that connects all the spectrum of the cloud-edge continuum, based on many factors (\eg latency, computation power) to ease the deployment and migration of \wasm applications is yet another milestone to reduce the gap between the cloud and the edge worlds.
We are confident that \wasm and trusted computing can serve as the foundation for software development for large-scale systems in the years to come.
 \paragraph*{Acknowledgements}
This publication incorporates results from the VEDLIoT project, which received funding from the European Union’s Horizon 2020 research and innovation programme under grant agreement No 957197.
 
\printbibliography

\end{document}